\documentclass[pre,aps,preprint,showpacs,preprintnumbers,amsmath,amssymb,secnumroman,eqsecnum]{revtex4}

\usepackage{graphicx}
\usepackage{dcolumn}
\usepackage{bm}

\newcommand{\beq}{\begin{equation}}
\newcommand{\eeq}{\end{equation}}
\newcommand{\beqa}{\begin{eqnarray}}
\newcommand{\eeqa}{\end{eqnarray}}

\newcommand{\vol}[1]{{\bf #1}}

\begin{document}


\title{Comment on: "Circular Motion of Asymmetric Self-Propelling Particles"}

\author{B. U. Felderhof}

 \email{ufelder@physik.rwth-aachen.de}
\affiliation{Institut f\"ur Theoretische Physik A\\ RWTH Aachen\\
Templergraben 55\\52056 Aachen\\ Germany\\
}%

\date{\today}

\begin{abstract}

\end{abstract}

\maketitle
In a recent Letter K\"ummel et al.\cite{1} have studied self-propulsion of an asymmetric particle in a viscous fluid at low Reynolds number both experimentally and theoretically. They claim full agreement between theory and experiment. However, the theory is conceptually wrong. If a particle exerts a net force, then it gives rise to a flow pattern which in infinite fluid falls off inversely with distance from the particle. If it exerts a net torque the flow pattern falls off inversely with distance squared. These long-range flow patterns are absent in self-propulsion.

We have remarked elsewhere\cite{2} that three different types of particle motion in a viscous fluid should be distinguished, namely pulling, self-propulsion, and swimming. In the case of pulling a steady external force and/or torque is applied to the body, resulting in a translational and/or rotational velocity. The relationship between force, torque, and the two velocities is linear and can be expressed in terms of a six-dimensional friction matrix, or its inverse, the mobility matrix\cite{3}.

In the case of self-propulsion the net force and torque exerted by the body on the fluid vanish. The body is rigid and acquires a translational and/or rotational velocity by means of a mechanism which creates a flow or pressure disturbance in its neighborhood. For example, the propulsive mechanism may involve chemical reactions in the fluid just outside the body. The disturbance acts back on the body, and fluid and body acquire equal and opposite momentum and angular momentum.

In the case of swimming the net force and torque exerted by the body on the fluid again vanish\cite{4},\cite{5}. The body acquires a velocity due to a surface distortion which varies in time, possibly in periodic fashion\cite{6}. The flow is caused by the no-slip boundary condition at the varying body surface.

The authors\cite{1} deal with self-propulsion, but talk also about microswimmers, even though the particle is rigid. Their theoretical calculation is based on the mobility matrix of the asymmetric particle. Since the mobility matrix applies to pulling, the calculation is inappropriate.

The calculation of the convective effect in self-propulsion can be based on Fax\'en's theorem\cite{3}. This yields the particle velocity in the presence of a flow field and/or pressure disturbance, in the absence of force or torque. We have performed such a calculation for a particular example of pressure disturbance due to chemical reactions near a spherical particle\cite{2}. Collisions with the fluid molecules may result in Brownian motion in addition to the systematic particle velocity calculated from the theorem. We have explained in a time-dependent picture\cite{2} how an impulsive event in the fluid can lead to a displacement of the particle. A random sequence of such events causes a net displacement with both a systematic and a stochastic component.

Unfortunately, for an asymmetric particle Fax\'en's theorem is not easily applied\cite{7}. For a proper calculation it would be necessary to know the Fax\'en relation for the given shape. Also the flow and pressure disturbance should be known for a definite prediction of the particle velocity.

\end{document}